\documentclass[showpacs,preprint,aps,prl]{revtex4}
\usepackage{epsfig}
\usepackage{epsf}
\usepackage{amsmath}
\usepackage{amssymb}
\usepackage{graphicx}

\begin{document}



\title{Entanglement of Accelerating Particles}

\author{W. L. Ku}
\author{M.-C. Chu}
\affiliation{   Department of Physics and Institute of Theoretical Physics, the Chinese University of Hong Kong, Hong Kong SAR, PRC}

\date{\today}

\begin{abstract}
 We study how the entanglement of a maximally entangled pair of particles is
 affected when one or both of the pair are uniformly accelerated, while the
 detector remains in an inertial frame. We find that the
 entanglement is unchanged if all degrees of freedom are considered.
 However, particle pairs are produced, and the entanglements of different bipartite
 systems may change with the acceleration.
 In particular,
 the entanglement between accelerating fermions is transferred
 preferentially to the produced antiparticles when the acceleration is large,
 and the entanglement transfer is complete when the acceleration approaches infinity.
 However, for scalar particles, no entanglement transfer to the antiparticles is observed.

\end{abstract}


\maketitle

  Entanglement is an important property of quantum mechanical systems.
  It is useful in the field of quantum information and quantum computing,
  such as in quantum teleportation~\cite{qutele}. It also
  finds many applications in quantum control~\cite{qc} and quantum
  simulations~\cite{qs}. Studying quantum entanglement in relativistic systems may give us
  insights on the relationship between quantum mechanics and general
  relativity. It has been shown that entanglement is
  Lorentz invariant~\cite{adami,alsing}. However, an accelerating observer
  measures less entanglement than an inertial observer in both the
  scalar~\cite{tele,nb} and fermion~\cite{nf} cases. This degradation in
  entanglement is due to the splitting of space-time, as a
  result of which the vacuum observed in one frame can become excited in
  another frame - the case of Unruh
  effect~\cite{unruh}. Classically, the trajectory of a uniformly
  accelerating particle observed by an inertial observer is the
  same as that of an inertial particle measured by a uniformly accelerating
  observer with appropriate acceleration, and it will be interesting to study how the acceleration of
  particles affects the entanglement of the originally entangled states. Pair production occurs
  when a relativistic particle accelerates. We will show that the entanglement remains unchanged if all
  degrees of freedom are considered. However, the entanglements between different bipartite systems may
  change with the acceleration. In particular, for fermions the entanglement is preferentially transferred
  to the antiparticles when the acceleration is large, and the entanglement transfer is complete when the
  acceleration approaches infinity. However, for scalar particles,  no entanglement transfer
  to the antiparticles is observed. Entanglement transfer to produced particles is a unique quantum mechanical
  and relativistic phenomenon, and it may play a role in many problems such as the black hole information paradox.

  We add an electric field in the Dirac and Klein-Gordon equations that would lead to uniform acceleration
  in the classical limit. A strong electric field makes the vacuum unstable and leads to pair
  production ~\cite{fs,wh,jsch}, which has been studied in the time
  dependent gauge~\cite{pptg_1,pptg_2}, in Rindler coordinates
  ~\cite{ppr} and in a finite region ~\cite{ppf_1,ppf_2,ppf_3}. We
  quantize the fields
  and use the in/out formalism to calculate the pair production,
  and we calculate the
  entanglements in different bipartite systems. The results are compared with those observed by an
  accelerating observer. Entanglements of different degrees of
  freedom will be shown.


The entanglement of a bipartite system can be quantified by the
logarithmic negativity~\cite{lne,ne_1,ne_2}. For a density operator
$\rho_{A,B}$ corresponding to a bipartite system $A$ and $B$, we
define the trace norm $||\rho_{A,B}||$ $\equiv$ $tr|\rho_{A,B}|$ =
$tr\sqrt{\rho_{A,B}^{\dagger}\rho_{A,B}}$, and the negativity,
$N_{e}\equiv({||\rho^{T_{A}}||-1})/{2}$, where $\rho^{T_{A}}$ is the
partial transpose of $\rho_{A,B}$ with respect to the party A.
$N_{e}$ can be calculated from the absolute value of the sum of the
negative eigenvalues of $\rho^{T_{A}}$. Then the logarithmic
negativity of the bipartite system $A$ and $B$ is defined by,

\begin{eqnarray}
LN(\rho_{A,B})\equiv\log_{2}||2N_{e}+1||.
\end{eqnarray}
For a product state, $LN(\rho_{A,B})$ = 0, and for entangled states,
$LN(\rho_{A,B})>0$.

The Klein-Gordon equation~$(\hbar=c=1)$ for a unit-charged particle
with mass $m$ in a uniform electric field $E$ ~\cite{kgwp_0,kgwp_1}
is

\begin{eqnarray}
(D_{\mu}D^{\mu}+m^{2})\phi =0,\label{okg}
\end{eqnarray}
where $D_{\mu}=\partial_{\mu}+iA_{\mu}$ and the gauge is chosen to
be $A_{0}$ = $-Ex$ and $A_{x}$ = 0. By assuming the form of solution
as, $\phi_{\omega}(t,x)=C\exp({i\omega t})\chi_{\omega}(x)$, where
$C$ is a normalization constant, we obtain from Eq.~(\ref{okg})

\begin{eqnarray}
\left[\frac{\partial^{2}}{\partial
x^{2}}+E^{2}(x-\omega/E)^{2}\right]\chi_{\omega}(x)=m^{2}\chi_{\omega}(x).
\label{simkg}
\end{eqnarray}
The solutions of Eq.~(\ref{simkg}) can be found in ~\cite{hand}, and
they are parabolic cylinder functions, $D_{-a-\frac{1}{2}}(x)$. We
can classify the solutions in the in/out basis
~\cite{kgwp_0,greiner} and have the in-basis functions,
\begin{eqnarray}
&&\phi^{in}_{\omega,p}(x,t)
=\frac{e^{-3\pi\mu^{2}/4}}{(2E)^{1/4}}e^{i\omega
t}D_{i\mu^{2}-1/2}[e^{-3i\pi/4}\sqrt{2E}(x-\omega/E)],
\\ &&
\phi^{in}_{\omega,a}(x,t)=\phi^{in}_{-\omega,p}(-x,t),
\end{eqnarray}
where $\mu^{2}=m^{2}/2E$. The subscripts $p$ and $a$ stand for
particles and antiparticles respectively. We also obtain the
out-basis solutions,

\begin{eqnarray}
&& \phi^{out}_{\omega,p}(x,t)=\phi^{in*}_{\omega,p}(x,-t),
\\ &&
\phi^{out}_{\omega,a}(x,t)=\phi^{in*}_{-\omega,p}(-x,-t).
\end{eqnarray}

As there are two different complete bases, we can quantize the field
in two ways,

\begin{eqnarray}
&&
\phi=\sum_{\omega}(a_{\omega}^{in}\phi_{\omega,p}^{in}+b_{\omega}^{in\dagger}\phi_{\omega,a}^{in*}),
\end{eqnarray}
or
\begin{eqnarray}
\phi=\sum_{\omega}(a_{\omega}^{out}\phi_{\omega,p}^{out}+b_{\omega}^{out\dagger}\phi_{\omega,a}^{out*}).
\end{eqnarray}
The operators $a_{\omega}^{in}$($b_{\omega}^{in\dagger}$),
$a_{\omega}^{out}$($b_{\omega}^{out\dagger}$) are the
annihilation~(creation) operators in the in-basis and out-basis, and
they are related by the Bogoliubov transformation,

\begin{eqnarray}
&& a_{\omega}^{out}=\alpha^{*}a^{in}_{\omega}-\beta^{*}
b_{\omega}^{in\dagger},
\\ && \nonumber
b_{\omega}^{out}=\alpha^{*}b_{\omega}^{in}-\beta^{*}
a_{\omega}^{in\dagger},
\end{eqnarray}
where

\begin{eqnarray}
&&\label{scboc}
\alpha=\frac{\sqrt{2\pi}e^{-i\pi/4}e^{-\pi\mu^{2}/2}}{\Gamma(1/2+i\mu^{2})},
\\ && \nonumber
\beta=e^{i\pi/2}e^{-\pi\mu^{2}},
\end{eqnarray}
which have the relation

\begin{eqnarray}
|\alpha|^{2}-|\beta|^{2}=1. \label{scarela}
\end{eqnarray}

We can express the in-vacuum state as a linear combination of out
states,
\begin{eqnarray}
|0\rangle_{in}=\prod_{\omega}\frac{1}{\alpha}\exp\left[\left(-\frac{\beta^{*}}{\alpha}\right)\sum_{\omega}a^{out\dagger}_{\omega}b^{out\dagger}_{\omega}\right]|0\rangle_{out}.
\end{eqnarray}
We let $\alpha=e^{i\phi_{1}}\cosh{r}$  and
$\beta=e^{i\phi_{2}}\sinh{r}$, where $r$ is a parameter related to
the acceleration (via Eq.~\ref{scboc}), and we ignore the phase
factors, which do not affect the following calculations of
entanglement. Taking the single-mode approximation, we get the
in-vacuum state in terms of the out states,

\begin{eqnarray}
&&
|0_{p}\rangle_{in}=\frac{1}{\cosh{r}}\sum_{n=0}^{\infty}\tanh^{n}{r}|n_{p}\rangle_{out}|n_{a}\rangle_{out}.
\end{eqnarray}
Similarly the one-particle state is
\begin{eqnarray}
|1_{p}\rangle_{in}
=\frac{1}{\cosh^{2}{r}}\sum^{\infty}_{n=0}\tanh^{n}{r}\sqrt{n+1}|(n+1)_{p}\rangle_{out}|n_{a}\rangle_{out}.
\end{eqnarray}

For a unit-charged fermions with mass $m$ coupled to an uniform
electric field,

\begin{eqnarray}
[\gamma^{\mu}\pi_{\mu}-m]\psi=0 \label{diraceqt},
\end{eqnarray}
where
\begin{eqnarray}
\pi_{\mu}\equiv i\partial_{\mu}-A_{\mu}.
\end{eqnarray}
Here, $A_{\mu}$ is the vector potential, and $\gamma_{\mu}$ is the
gamma matrix. Then we let
\begin{eqnarray}
\psi=(\gamma^{\nu}\pi_{\nu}+m)\phi \label{diraclet}
\end{eqnarray}
to obtain
\begin{eqnarray}
\label{diracsim}
\left[\pi^{2}-m^{2}-\frac{i}{2}\gamma^{\mu}\gamma^{\nu}(\partial_{\mu}A_{\nu}-\partial_{\nu}A_{\mu})\right]\phi=0.
\end{eqnarray}
We choose the gauge to be $A_{0}=0$ , $A_{3}=-Et$  and get back a
Klein-Gordon equation~\cite{ppf_1} from Eq.~(\ref{diracsim}). The
solution is still the parabolic cylinder function. The in/out basis
solution of the second order ODE is still the in/out basis solution
of the Dirac equation Eq.~(\ref{diraceqt}). Therefore, we obtain the
Bogoliubov coefficients, which have been calculated in
Ref.~\cite{niki},

\begin{eqnarray}
&& a_{n}^{out}=\alpha_{f} a_{n}^{in}-\beta_{f}^{*}b_{n}^{in\dagger},
\\ && \nonumber
b_{n}^{out\dagger}=\beta_{f} a_{n}^{in}+\alpha_{f}^{*}b_{n}^{in\dagger},
\end{eqnarray}
where

\begin{eqnarray}
\label{ferminout}&&\beta_{f}=e^{-\pi\mu^{2}},
\\ && \nonumber
\alpha_{f}^{*}=-i\sqrt{\frac{2\pi}{\mu^{2}}}\frac{e^{-\pi\mu^{2}/2}}{\Gamma(i\mu^{2})},
\end{eqnarray}
with $\alpha_{f}$ and $\beta_{f}$ having the relation,

\begin{eqnarray}
|\alpha_{f}|^{2}+|\beta_{f}|^{2}=1. \label{fermrela}
\end{eqnarray}

We let $\alpha_{f}$ = $\cos{r_{f}}e^{i\phi}$ and $\beta_{f}$ =
$\sin{r_{f}}$, $r_{f}$ being a parameter with values between 0 and
${\pi}/{2}$ and related to the acceleration (via
Eq.~\ref{ferminout}). Also, we can relate the incoming states with
the outgoing states as in the case of an accelerating
detector~\cite{nf},

\begin{eqnarray}
\label{fermint}&|0_{p}\rangle_{in}&
=\cos{r_{f}}e^{-i\phi}|0_{p}\rangle_{out}|0_{a}\rangle_{out}-
\sin{r_{f}}|1_{p}\rangle_{out}|1_{a}\rangle_{out},
\\  \nonumber
&|1_{p}\rangle_{in}&=|1_{p}\rangle_{out}|0_{a}\rangle_{out}.
\end{eqnarray}

Initially, we have the incoming entangled state,
\begin{eqnarray}
\label{ferins}\Psi_{i}=\frac{1}{\sqrt{2}}\left[|0_{s,p}\rangle_{in}|0_{\omega,p}\rangle_{in}+
|1_{s,p}\rangle_{in}|1_{\omega,p}\rangle_{in}\right].
\end{eqnarray}
Then either one or both of the particles in $\omega$ and $s$ modes
are accelerated by the electric field, and the in states in
Eq.~(\ref{ferins}) are replaced by the out states in
Eq.~(\ref{fermint}). If only the $\omega$ mode is accelerated, we
have

\begin{flushleft}
\begin{eqnarray}
\Psi_{f}  &=& \frac{1}{\sqrt{2}}\left\{|0_{s,p}\rangle_{out} \right.
 \otimes\left[\cos{r_{f}}e^{-i\phi}|0_{\omega,p}\rangle_{out}|0_{\omega,a}\rangle_{out}-
\sin{r_{f}}|1_{\omega,p}\rangle_{out}|1_{\omega,a}\rangle_{out}\right]
\\ && \nonumber \left.
+|1_{s,p}\rangle_{out}\otimes(|1_{\omega,p}\rangle_{out}|0_{\omega,a}\rangle_{out})\right\}.
\end{eqnarray}
\end{flushleft}
If both the $s$ and $\omega$ modes are accelerated with the same $r_{f}$, we have
\begin{flushleft}
\begin{eqnarray}
\Psi_{f}
&=&\frac{1}{\sqrt{2}}\left\{\left[\cos{r_{f}}e^{-i\phi_{1}}|0_{s,p}\rangle_{out}|0_{s,a}\rangle_{out}
 \right.
-\sin{r_{f}}|1_{s,p}\rangle_{out}|1_{s,a}\rangle_{out}\right]
\\ && \nonumber
\otimes\left[\cos{r_{f}}e^{-i\phi_{1}}|0_{\omega,p}\rangle_{out}|0_{\omega,a}\rangle_{out}
-\sin{r_{f}}|1_{\omega,p}\rangle_{out}|1_{\omega,a}\rangle_{out}\right]
\\ && \nonumber \left.
+\left[(|1_{s,p}\rangle_{out}|0_{s,a}\rangle_{out})\otimes(|1_{\omega,p}\rangle_{out}|0_{\omega,a}\rangle_{out})\right]\right\}.
\end{eqnarray}
\end{flushleft}

The degradation of entanglement in the case of an accelerating
detector is due to the fact that some degrees of freedom have been
traced out. An accelerating detector 'sees' the space-time being
split into two causally disconnected regions, and it cannot access
information in one of them. We have verified explicitly that the
entanglement between the particles in $s$ mode and $\omega$ mode is
unchanged if there is no tracing out of any space-time region. On
the other hand, in the case of accelerating particles, the detector,
which is in an inertial frame, can access all degrees of freedom and
the orthogonality of the states is unchanged; therefore, the
entanglement of accelerating particles is unchanged.

However, more degrees of freedom are produced and we can calculate
the entanglements between different bipartite systems. In
Ref.~\cite{adami}, it was shown that entanglement is Lorentz
invariant. If one traces out the momentum, the entanglement
decreases, and the entanglement is transferred between the momentum
and the spin degrees of freedom. We will show that entanglement
transfer also occurs in accelerating fermions. Detailed calculations
for both non-relativistic and relativistic particles are shown in
\cite{longer}.

\begin{figure}
\includegraphics{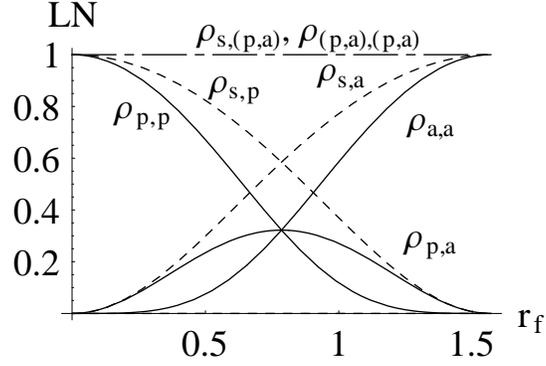}
\caption{\label{fig:bfacc} Logarithmic negativities of several
bipartite systems when one or both fermions are accelerated, the
magnitude of which is parameterized by $r_{f}$. In both cases, the
entanglement between the entire $s$ mode and $\omega$ mode is
unchanged (dot-dashed line). The solid lines show the results when
both particles are accelerated together, for three bipartite
systems: particles in $s$ mode and particles in $\omega$ mode
($\rho_{p,p}$), particles in $s$ mode and antiparticles in $\omega$
mode ($\rho_{p,a}$), and antiparticles in $s$ mode and antiparticles
in $\omega$ mode ($\rho_{a,a}$).  For comparison, the dashed lines
show the results when only the particle in the $\omega$ mode is
accelerated, in which case the two bipartite systems are particle in
$s$ mode and particles in $\omega$ ($\rho_{s,p}$), and particle in
$s$ mode and antiparticles in $\omega$ ($\rho_{s,a}$). }
\end{figure}

If only the particle in the $\omega$ mode is accelerated, we can
study the three bipartite systems: $A$ = the $s$ mode, $B$ = the
particles in the $w$ mode, the antiparticles in $w$ mode, or the
entire $w$ mode including both the particles and antiparticles. The
density matrices are called $\rho_{s,p}$, $\rho_{s,a}$, and
$\rho_{s,(p,a)}$ respectively. The entanglements are
\begin{eqnarray}
\left\{
  \begin{array}{ll}
    LN(\rho_{s,(p,a)})=1, &\ \\
    LN(\rho_{s,p})=\textrm{log}_{2}(1+\cos^{2}{r_{f}}), & \ \\
    LN(\rho_{s,a})=\textrm{log}_{2}(1+\sin^{2}{r_{f}}), &\
  \end{array}
\right.
\end{eqnarray}
which are plotted in Fig.~\ref{fig:bfacc}. It is obvious that the
entanglement of $\rho_{s,p}$ is transferred to $\rho_{s,a}$.

When both the $s$ and $\omega$ modes are accelerated with the same
$r_{f}$, we can calculate the entanglements between the five
bipartite systems: particles in $s$ mode and particles in $\omega$
mode ($\rho_{p,p}$), antiparticles in $s$ and antiparticles in
$\omega$ ($\rho_{a,a}$), antiparticles in $s$ and particles in
$\omega$ ($\rho_{a,p}$), particles in $s$ and antiparticles in
$\omega$ ($\rho_{p,a}$), and the entire $s$ and $\omega$ modes
($\rho_{(p,a),(p,a)}$). The logarithmic negativities are
\begin{eqnarray}
\left\{
  \begin{array}{ll}
    LN(\rho_{(p,a),(p,a)})=1,  \\
    LN(\rho_{p,p})=\textrm{log}_{2}\left[1+\cos^{4}{r_{f}}\right],  \\
    LN(\rho_{a,a})=\textrm{log}_{2}\left[1+\sin^{4}{r_{f}}\right],  \\
     LN(\rho_{p,a})=\textrm{log}_{2}\left[1+\cos^{2}{r_{f}}\sin^{2}{r_{f}}\right].
  \end{array}
\right.
\end{eqnarray}
By symmetry, $LN (\rho_{a,p}) = LN(\rho_{p,a})$. The results are
shown in Fig.~\ref{fig:bfacc}. The entanglement is transferred from
$\rho_{p,p}$ not only to $\rho_{p,a}$, but also to $\rho_{a,a}$. In
fact, when the acceleration of the particles tends to infinity, the
entanglement is completely transferred to between the antiparticles
$\rho_{a,a}$.

\begin{figure}
\includegraphics{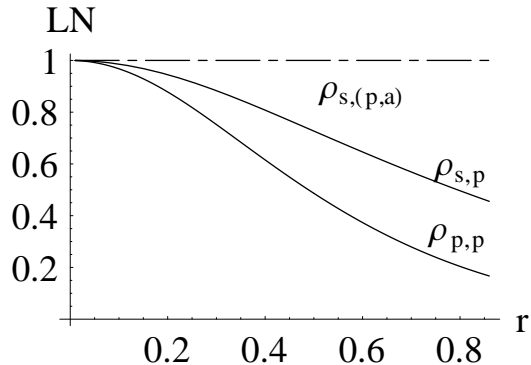}
\caption{\label{fig:bsacc_all} Same as Fig.~\ref{fig:bfacc}, but for
scalar particles.  Again, the entanglement of the entire $s$ mode
and $\omega$ mode~(both particles and antiparticles), indicated by
the dot-dashed line, remains unchanged. Note that $LN(\rho_{s,a}) =
LN(\rho_{p,a}) = LN(\rho_{a,a}) = 0$ for all $r$. }
\end{figure}

Similarly, for scalar particles, we calculate the entanglements of $\rho_{s,p}$, $\rho_{s,a}$,
$\rho_{p,p}$, $\rho_{a,a}$ and $\rho_{p,a}$. The results are shown
in Fig.~\ref{fig:bsacc_all}.
Again, the entanglement between the entire $s$ and $\omega$ modes
remains unchanged, such that $LN(\rho_{s,\omega}) = LN(\rho_{(p,a),
(p,a)}) = 1$ for all $r$.  In contrast to fermions, there is no
entanglement transfer to the antiparticles, and $LN(\rho_{s,a}) =
LN(\rho_{p,a}) = LN(\rho_{a,a}) = 0$ for all $r$, even though the
entanglement between the particles in the $s$ and $\omega$ modes
decreases as $r$ increases.

We have studied how the entanglement of a pair of maximally
entangled particles is affected when one or both of the pair is
uniformly accelerated, as measured by an inertial detector, and
compared it with that of inertial particles observed by a uniformly
accelerating detector. While there is a degradation of entanglement
in the latter case due to the splitting of the space-time, the
entanglement in the former case is unchanged by the acceleration
when all degrees of freedom are considered. However, particle pairs
are produced, and the entanglements of different bipartite systems
may change as the acceleration. In particular, for fermions, the
entanglement is preferentially transferred to the produced
antiparticles and when the acceleration approaches infinity, the
entanglement is completely transferred to the antiparticles.
However, for scalar particles, no entanglement transfer to the
antiparticles is observed.

Our results raise the possibility that when an entangled pair falls
into a black hole, their entanglement may be partially transferred
to the produced particles, which should not be ignored in
considering the black hole information paradox. Studying quantum
entanglement in curved space-time may therefore give us insights on
the relation between quantum mechanics and general relativity.


\end{document}